\documentclass{emulateapj}

\usepackage{macros}

\shorttitle{RRLs toward GCL}
\shortauthors{Law et al.}

\begin{document}


\title{Radio Recombination Lines toward the Galactic Center Lobe}

\author{C. J. Law\altaffilmark{1,2,3}}
\author{D. Backer\altaffilmark{3}}
\author{F. Yusef-Zadeh\altaffilmark{1}}
\author{R. Maddalena\altaffilmark{4}}

\altaffiltext{1}{Department of Physics and Astronomy, Northwestern University, Evanston, IL 60208, USA; claw@astro.berkeley.edu, zadeh@northwestern.edu}
\altaffiltext{2}{Astronomical Institute ``Anton Pannekoek'', University of Amsterdam, Kruislaan 403, 1098 SJ Amsterdam, The Netherlands}
\altaffiltext{3}{Radio Astronomy Lab, University of California, Berkeley, CA 94720, USA; dbacker@astro.berkeley.edu}
\altaffiltext{4}{National Radio Astronomy Observatory, Green Bank, WV, 24944, USA; rmaddale@nrao.edu}

\begin{abstract}
The Galactic Center lobe is a degree-tall shell seen in radio continuum images of the Galactic center (GC) region.  If it is actually located in the GC region, formation models would require massive energy input (e.g., starburst or jet) to create it.  At present, observations have not strongly constrained the location or physical conditions of the GC lobe.  This paper describes the analysis of new and archival single-dish observations of radio recombination lines toward this enigmatic object.  The observations find that the ionized gas has a morphology similar to the radio continuum emission, suggesting that they are associated.  We study averages of several transitions from H106$\alpha$ to H191$\epsilon$ and find that the line ratios are most consistent with gas in local thermodynamic equilibrium.  The radio recombination line widths are remarkably narrow, constraining the typical electron temperature to be less than about 4000 K.  These observations also find evidence of pressure broadening in the higher electronic states, implying a gas density of $n_e=910^{+310}_{-450}$ cm$^{-3}$.  The electron temperature, gas pressure, and morphology are all consistent with the idea that the GC lobe is located in the GC region.  If so, the ionized gas appears to form a shell surrounding the central 100 parsecs of the galaxy with a mass of roughly $10^5$ \msol, similar to ionized outflows seen in dwarf starbursts.
\end{abstract}


\section{Introduction}
Since their prediction and discovery more than 40 years ago \citep{k59,d64,s64}, radio recombination lines have been useful tools for probing the physical conditions of the ISM.  The properties of thermal emission can constrain fundamental parameters such as gas temperature and density \citep[e.g.,][]{r92}.  Furthermore, the spectral line can provide valuable information about the kinematics of the emitting gas.

In the Galactic Center (GC) region, there have been extensive observations of radio recombination lines.  These lines are useful because they are detectable despite heavy extinction and can be used to separate thermal \hii\ regions from the widespread nonthermal gas \citep{p75}.  Separating thermal and nonthermal emission is especially important in the GC region, which is filled with extended ionized gas and synchrotron-emitting features \citep[e.g.,][]{y84,m96,l97}.  Recombination line surveys have found that diffuse ionized gas is strongly correlated with the molecular gas density and star formation, suggesting that the GC region should be filled with ionized gas \citep{l96}.  This gas can provide a direct probe of the conditions in the GC region.  

The information available by studying ionized gas could help address questions surrounding a feature in the GC region known as the Galactic center lobe (GCL).  The GCL was first noted in a 10 GHz radio continuum survey of the central few degrees of the Galaxy \citep{s84}.  The survey found a degree-tall loop of emission north of the galactic plane, surrounding the central degree, covering an area from $l=359\ddeg2-0\ddeg2$ and $b=0\ddeg2-1\ddeg2$.  The resemblance of the GCL to a wind-driven mass outflow was striking and led to several models for its formation \citep[e.g.,][]{c92,m01}.  However, while the first study of the radio spectral index found it to be consistent with thermal emission \citep{s85}, later work suggested the GCL was nonthermal \citep{t86,r87}.  \citet{b03} noted the morphological connection between the radio continuum and mid-IR emission and proposed a starburst outflow model.  However, the connection of the ionized gas to this feature is still unclear.

Motivated by a desire to understand the nature of the emission in the GCL, we have conducted new radio recombination line observations and collected archived observations in this region.  This paper describes observations of the GCL with the Robert C. Byrd 100-m Green Bank Telescope\footnote{The National Radio Astronomy Observatory is a facility of the National Science Foundation operated under cooperative agreement by Associated Universities, Inc.} (GBT) and the Hat Creek Radio Observatory 26-m telescope (HCRO).  Section \ref{recomb_obs} shows how the observations were made and calibrated.  Section \ref{recomb_res} describes the results of the observations, including the morphology, line characteristics, and kinematics of the line emission in the region.  The HCRO observations show that the H109$\alpha$ emission has a morphology similar to the radio continuum emission.  The GBT observations find that the recombination line emission is this region is in local thermodynamic equilibrium (LTE).  The unusually narrow recombination line emission throughout the GCL places useful constraints on the gas temperature and other properties.  In \S\ \ref{recomb_dis}, we discuss the derived quantities of the ionized gas, including gas density, temperature, and the total mass (assuming a geometry and distance).  Section \ref{conc} summarizes the results of this paper.  A detailed discussion of the nature of the GCL is left for a later paper (Law, in prep).

\section{Observations and Data Reductions}
\label{recomb_obs}
\subsection{HCRO}
In 1985, we used the HCRO 26 m telescope to map the H109$\alpha$ recombination line toward the GCL.  At 5 GHz, the HCRO beam size is 10\arcmin\ and its Nyquist sample size is 3\damin3.  Pointings were made over a roughly 1\sdeg\ square region on a 6\arcmin\ grid;  neighboring points in the grid were skipped to form a checkerboard pattern.  Spectra were calibrated by position switching to a distant, fixed position, with a typical integration time of 2 hours for each spectrum.  Several scattered points outside the radio continuum emission associated with the GCL were also observed to constrain the background emission.  A crude gain calibration was made by comparing the brightness temperatures between the HCRO and GBT observations described below;  this comparison implies a gain of $\sim13$ Jy K$^{-1}$.

The spectrometer at HCRO was tuned to a central frequency of 5008.923 MHz, 512 channels, and a 20 MHz bandwith.  This bandwidth covered the H109$\alpha$ and H137$\beta$ transitions, although only the H109$\alpha$ transition was studied.  Spectra were Hanning smoothed and a seventh-order baseline fit before Gaussian line fitting.  The velocity uncertainty of the fits are about 1 \kms.  The system temperature was assumed to be 40 K, based on observations of empty sky.

Figure \ref{hcroimg} shows images of the H109$\alpha$ line antenna temperature and velocity from the HCRO data.  The line brightness is not corrected for atmospheric absorption.  Since the checkerboard observing pattern makes visualization awkward, the data were interpolated onto a regular grid.  The algorithm calculates a linear interpolation for missing pixels with four neighbors, then three, then two.  The order in which this interpolation is done creates some bias in the interpolated pixels, but the general properties of the emission on spatial scales larger than the 10\arcmin-beam are not affected.

\subsection{GBT}
\begin{figure}[tbp]
\includegraphics[width=\textwidth]{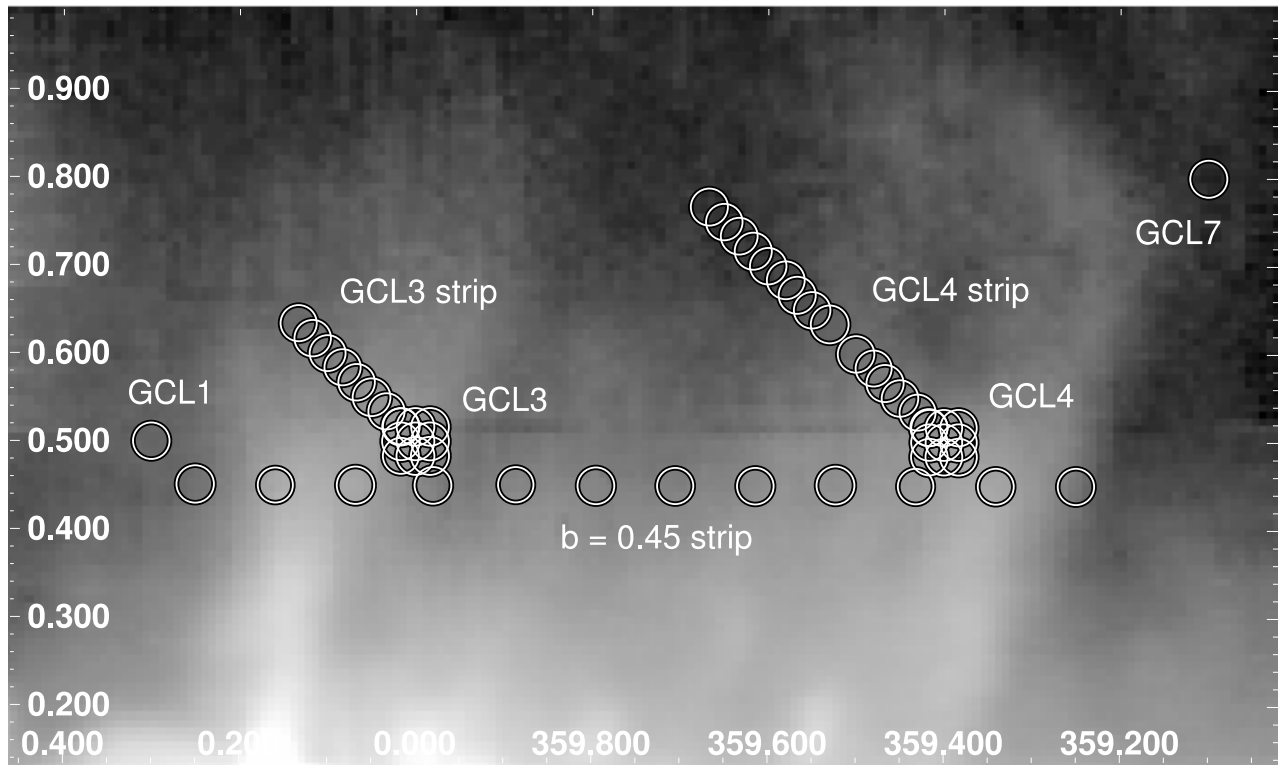}
\caption{Positions of the GBT observations toward the GCL are shown as circles on a GBT 5 GHz radio continuum survey of the region \citep{gcsurvey_gbt}.  The circles have a diameter of 2\damin5, which is the FWHM of the GBT beam at 5 GHz. \label{positions}}
\end{figure}

The goal of the GBT observation was to confirm the results of the HCRO observations and study the gas conditions in more detail.  We used the GBT Spectrometer in a 7 hour session in August 2005.  The GBT observations were more limited in spatial coverage than the HCRO survey; the locations are shown in Figure \ref{positions} and described in Table \ref{gbtpointings}.  The locations named ``GCL3'' and ``GCL4'' are at the two bright peaks in the HCRO map of H109$\alpha$.  These regions were mapped in a 3$\times$3, Nyquist-sampled (1\arcmin\ near 5 GHz) grid, with on-source integration times of 90 s per pointing.  Diagonal strips were observed starting at these positions and going to the Galactic northeast to study gradients in the velocity field.  A sparse horizontal strip was observed at $b=0\ddeg45$, to sample the changes in the east-west direction.  Finally, two positions outside the GCL, ``GCL1'' and ``GCL7'', were observed to measure background line emission.  The strips and individual pointings had integration times of 60 s per pointing.

The spectrometer was configured with four windows of width 200 MHz observing eight H$\alpha$ and He$\alpha$ transitions from $n=$106 to 113.   Each pointing produced dual-circular polarizations of four 8196 channel spectra.  The spectral windows, centered near 5.37, 5.08, 4.81, and 4.56 GHz, also covered other transitions, including:  H$\beta$ (7 transitions with $n=$134--142), H$\gamma$ (8; $n=$152--162), H$\delta$ (8; $n=$167--178), and H$\epsilon$ (8; $n=$180--191).  The spectra had 24.4 kHz channels, equivalent to 1.5 \kms\ at 4.9 GHz.  All velocities in this work are given in the local standard of rest (LSR).  Calibration for each scan was done by position switching to a position two degrees north in Galactic latitude.  

Calibration and analysis was done using tools released with GBTIDL v1.2.1 and custom IDL programs.  Spectra were calibrated using the standard (on-off)/off method.  The typical line analysis would apply standard calibration, average both polarizations and lines of the same transition (e.g., all H$\beta$ transitions;  rest frequencies from \citet{l68}), subtract a third order polynomial baseline fit, and fit a Gaussian line profile.  It is worth noting that the standard error propagation used does not account for the uncertainty in fitting a baseline to the spectra, and are likely to underestimate the true errors slightly.  By fitting with various assumptions, we estimated the baseline-induced uncertainty in the line fit is always less than the standard statistical errors in the line fit and generally decreases with signal to noise ratio of the line.

All spectra presented here are corrected for the beam efficiency and atmospheric opacity, and thus are in units of main beam brightness temperature.  The beam efficiency is assumed to be equal to 89\%. \footnote{See \url{http://wwwlocal.gb.nrao.edu/gbtprops/man/GBTpg/GBTpg\_tf.html}.}  This work corrects for mean atmospheric absorption opacity of 8.6$\times10^{-3}$, giving a correction factor of 2--5\% for these observations.

The GC region is seen at low elevations from Green Bank, which makes the calibration of the continuum levels difficult.  Unfortunately, the standard continuum calibration method could not properly account for atmospheric emission.  Since the observations described in \citet{gcsurvey_gbt} were made with a calibration method optimized for the study of continuum levels, we use results from that observation when continuum fluxes are needed.

\section{Results}
\label{recomb_res}

\subsection{HCRO}
\begin{figure}[tbp]
\includegraphics[width=\textwidth]{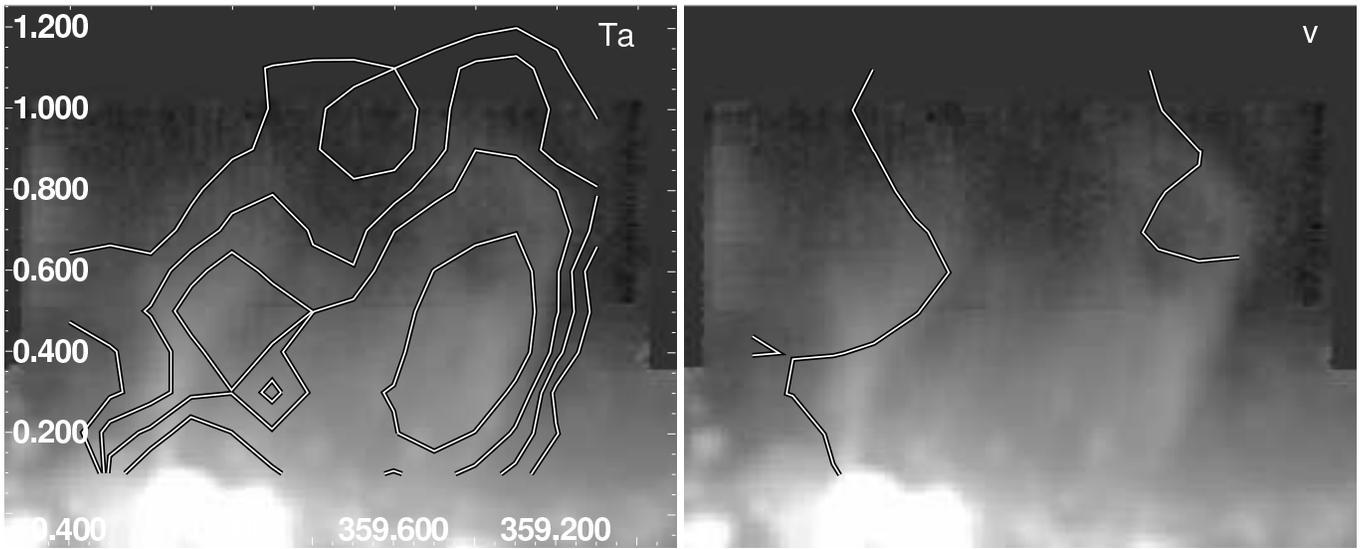}
\caption{\emph{Left}: The gray scale shows GBT 5 GHz radio continuum emission toward the GCL with contours of H109$\alpha$ brightness from the HCRO at $T_{\rm{a}}=$15, 20, 30, and 40 mK. \emph{Right}: Same as the left panel, but with a contour of H109$\alpha$ line velocity at 0 \kms.  The areas to the far west and east have negative velocities and between the lines are generally positive velocities. \label{hcroimg}}
\end{figure}

Figure \ref{hcroimg} shows the HCRO line intensity and velocity compared to 5 GHz continuum emission.  The radio recombination line is brightest along two vertical ridges near ($l, b$)=(0\ddeg0, 0\ddeg5) and (359\ddeg4, 0\ddeg5).  There is also a slight increase in brightness at the top of the line-emitting region, near ($l, b$)=(359\ddeg6, 1\ddeg0), like a cap to the structure.  The longitude of the line emission north of the plane is centered near $l=359\ddeg6$ and the central longitude shifts slightly to the west at higher latitudes.  The overall morphology of bright limbs with a cap mirrors that of the radio continuum emission.

The right side of Figure \ref{hcroimg} shows the best-fit velocity of the H109$\alpha$ line.  Two striking characteristics of the line velocities in this region are their small values and the absence of any simple pattern or gradient.  North of $b=0\ddeg3$ the line velocity does not exceed the range of -5 to +5 \kms.  This is consistent with previous observations of radio recombination lines in the region \citep{l73,p75,a97}.

There is significant emission at every HCRO pointing between the two ridges of emission.  At the lowest, the line brightness ranges from 10--30 mK, while the brightest line emission is 72 mK, at the peak of the western ridge.  The emission outside of the ridges of emission is generally less than 10 mK.  Detailed analysis of the gas properties depends on parameters that are not well constrained by the HCRO data.  The GBT results are more constraining and are consistent with the HCRO results, so the results are given below in \S\ \ref{gbt:lines}.

\subsection{GBT}
\label{recomb_gbt}

\subsubsection{Morphology}
\label{gbt:morph}
\begin{figure}[tbp]
\includegraphics[angle=270,scale=0.6]{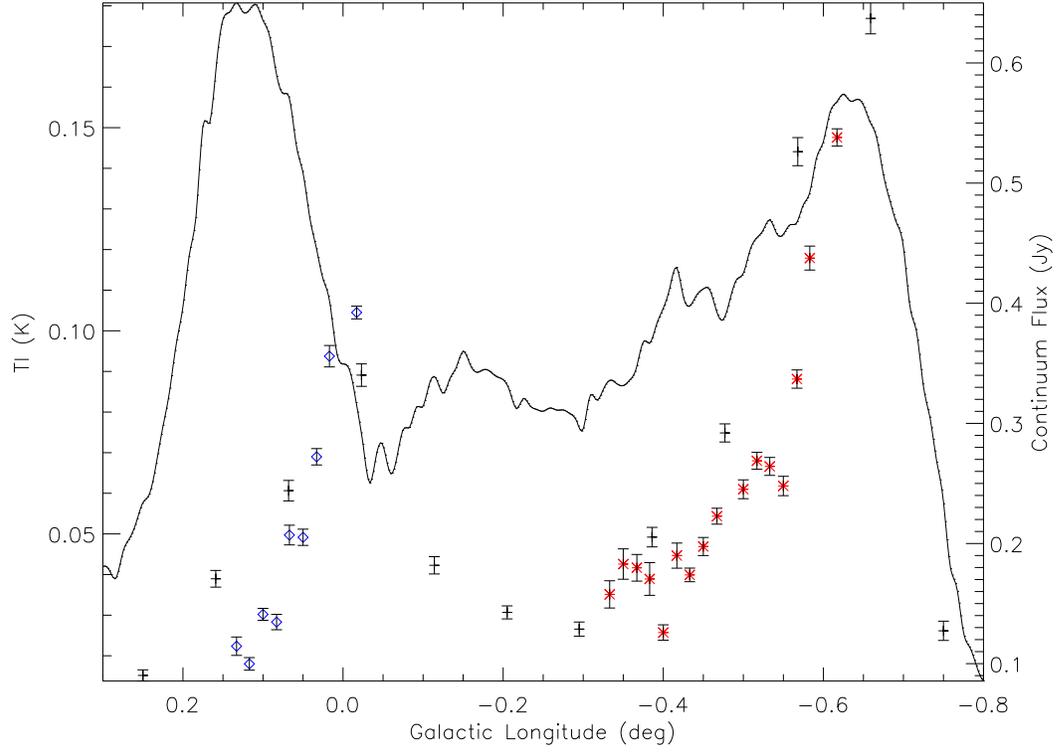}
\caption{Peak brightness temperature of H$\alpha$ as a function of Galactic longitude for three strips of GBT observations through the GCL.  The $b=0.45$ strip is shown with black crosses, diagonal ``GCL 4 strip'' shown with red stars, and diagonal ``GCL 3 strip'' shown with blue diamonds (see Table \ref{gbtpointings} for strip specifications).  Error bars show 1 $\sigma$ uncertainties.  The line shows the GBT 5 GHz continuum flux described in \citet{gcsurvey_gbt}, which defines the extent of the radio continuum GCL. \label{tastrips}}
\end{figure}

Figure \ref{tastrips} shows plots of the average H$106-113\alpha$ line brightness for the three strips of observations.  The horizontal strip at $b=0\ddeg45$ is similar to the HCRO data, with two recombination line peaks near the radio continuum peaks.  The eastern peak of the line emission is offset west the continuum peak by roughly $0\ddeg15$.  The western peak of the line and continuum emission fall at the same position ($l\sim359.35-359.4$), within the 5\arcmin\ spatial sampling of the line observations.  This is consistent with the morphology observed by the HCRO and again shows how the structure is similar to the radio continuum emission.

As with the HCRO observations, the line brightness between the two emission peaks is significantly brighter than outside of it.  The GCL1 and GCL7 pointings sampled the emission toward lines of sight outside the radio continuum emission of the GCL.  The best-fit Gaussian to the average H$106-113\alpha$ line in the GCL1 position had $(T_l, v, \Delta v)=(0.0103\pm0.0016\ \rm{K}, -9.5\pm2.2$\ \kms$, 29.2\pm5.2$\ \kms$)$.  No H$106-113\alpha$ line was detected toward GCL7 with a 1 $\sigma$ upper limit of 5.9 mK.  The line brightness for the weakest lines associated with the GCL have $T_l\sim25$\ mK.

The GCL3 and GCL4 diagonal strips of pointings, like the $b=0\ddeg45$ strip, show a decrease in peak line brightness as they move away from the two line emission peaks.  The longitude half-peak width of the east and west peaks are roughly $0\ddeg2$.

\subsubsection{Velocity Structure}
\label{gbt:vel}
\begin{figure}[tbp]
\begin{center}
\includegraphics[angle=270,scale=0.7]{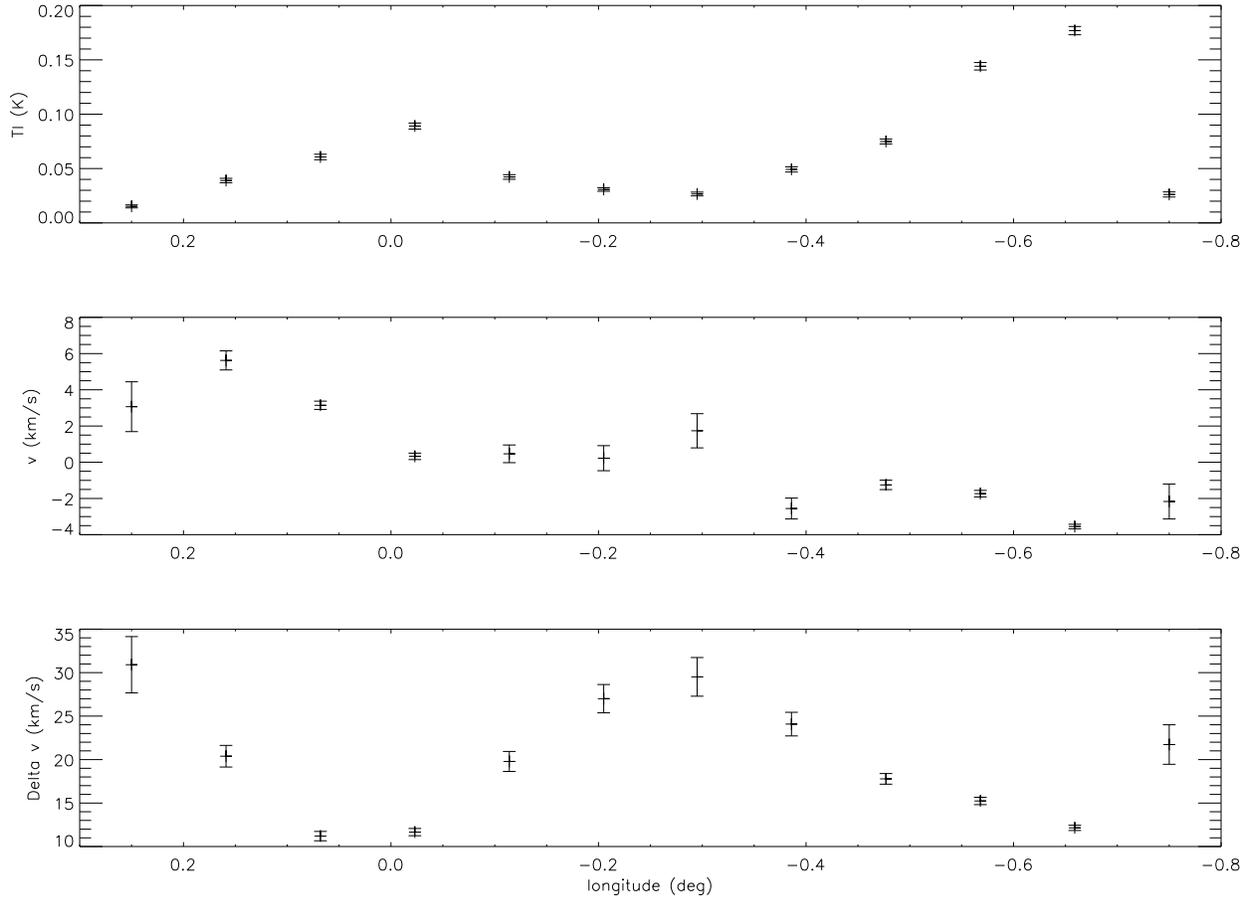}
\end{center}
\caption{Plots of the properties of the average H$\alpha$ recombination line for the series of pointings across the GCL at $b=0\ddeg45$.  Line properties found by fitting a single Gaussian to the profile.  The top plot shows the peak line brightness temperature, the middle plot shows the line velocity, and the bottom plot shows the line width.  The brightest recombination line emission (near $l\sim0$\sdeg\ and $-0\ddeg6$) has the narrowest line width. \label{polstart}}
\end{figure}

At the brightest parts of the recombination line emission, the line velocities observed by the GBT are consistent with those of the HCRO.  Figure \ref{polstart} shows how the average H$106-113\alpha$ line properties change as a function of Galactic longitude for the strip at $b=0\ddeg45$.  One interesting trend in Figure \ref{polstart} is the east-west asymmetry in the line velocities, with positive velocities on the east side and negative on the west.  Averaging over the four easternmost and westernmost scans in the $b=0\ddeg45$ strip gives spectra with a single Gaussian component with mean velocities of $2.3\pm0.2$ and $-2.4\pm0.2$ \kms, respectively.  This velocity gradient is consistent with the upper limit provided by the HCRO observations.

Comparing the peak line brightness and line widths in Figure \ref{polstart} shows an inverse relationship:  the brightest line emission has the narrowest widths and vice versa.  The top panel shows the line is brightest near $l=0$\sdeg\ and $l=-0\ddeg7$;  the bottom panel shows that the line is narrowest at these positions.  Since the line width is a useful constraint to the gas properties, we needed to understand the origin of this correlation.  

\begin{figure}[tbp]
\includegraphics[scale=0.15]{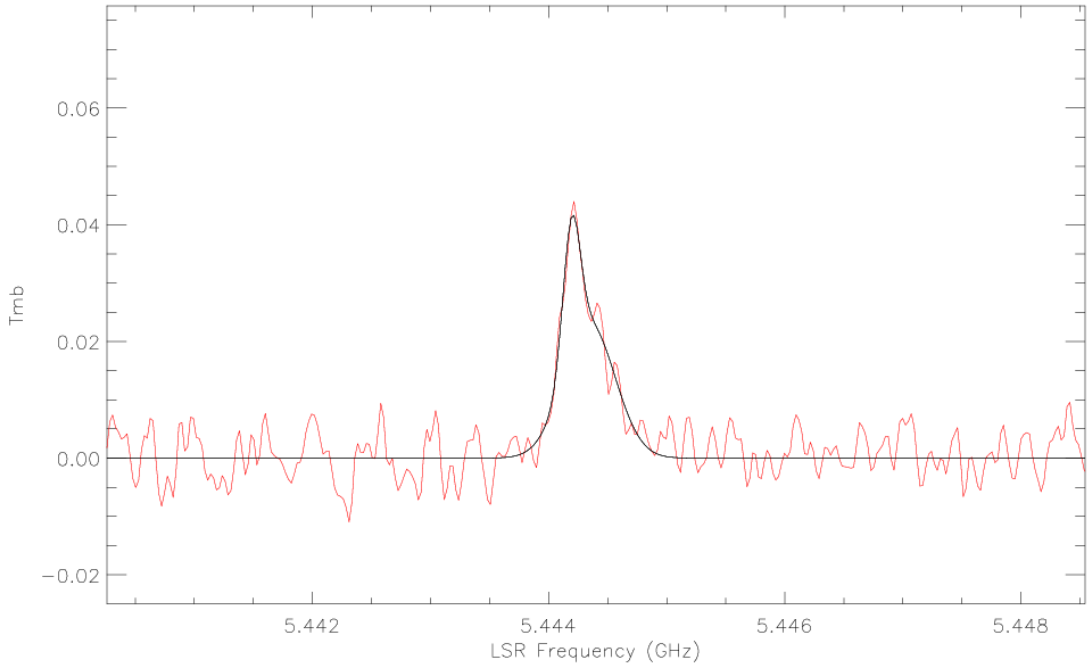}
\caption{Average H$\alpha$ spectrum for four scans near the center of the GCL with a line showing the best-fit model described in the text.  Fitting one Gaussian gives a moderately wide fit, but fitting two Gaussians gives a narrow and wide component of similar amplitude.  This shows how the assumption of a single Gaussian component can lead to a inverse relationship between line width and amplitude and that the unusually narrow line is observed throughout the GCL. \label{polcen}}
\end{figure}

Two potential explanations for the correlation are that (1) the line width is intrinsically tied to its brightness, or (2) the line has two components that can affect the best-fit line width according to their relative strengths.  Figure \ref{polcen} shows the average H$106-113\alpha$ profile for four scans located in the center of the GCL that, separately, have best-fit line widths of 20--30 \kms.  The average spectrum shows a slightly more complex line profile than seen near the brightest recombination line emission.  If the profile is fit with two Gaussians, their best-fit components are $(T_l, v, \Delta v)=(22\pm3\ \rm{mK}, 5.2\pm0.5\ $\kms$, 9.3\pm1.5\ $\kms$)$ and $(25\pm2\ \rm{mK}, -5.0\pm1.3$\ \kms$, 30.6\pm1.9$\ \kms$)$.  The narrow component is similar to that observed near the bright ridges of line emission, but its amplitude is comparable to the wide component, such that a single-Gaussian fit is moderately wide.  In averages of other scans, there is a tendency for a similar, wide (20--30 \kms), low-level (10--20 mK) line to appear.  The amplitude of this line is larger than the $\sim10$ mK seen in the background outside the GCL, so the wide line seems to be associated with the GCL.  Regardless of its origin, it seems that the changes in best-fit, single-Gaussian line width shown in Figure \ref{polstart} are likely due to the relative strength of the narrow line and this wider component.  Thus, there seems to be narrow ($9-14$ \kms) recombination line emission for all lines of sight through the GCL.

\begin{figure}[tbp]
\includegraphics[scale=0.6]{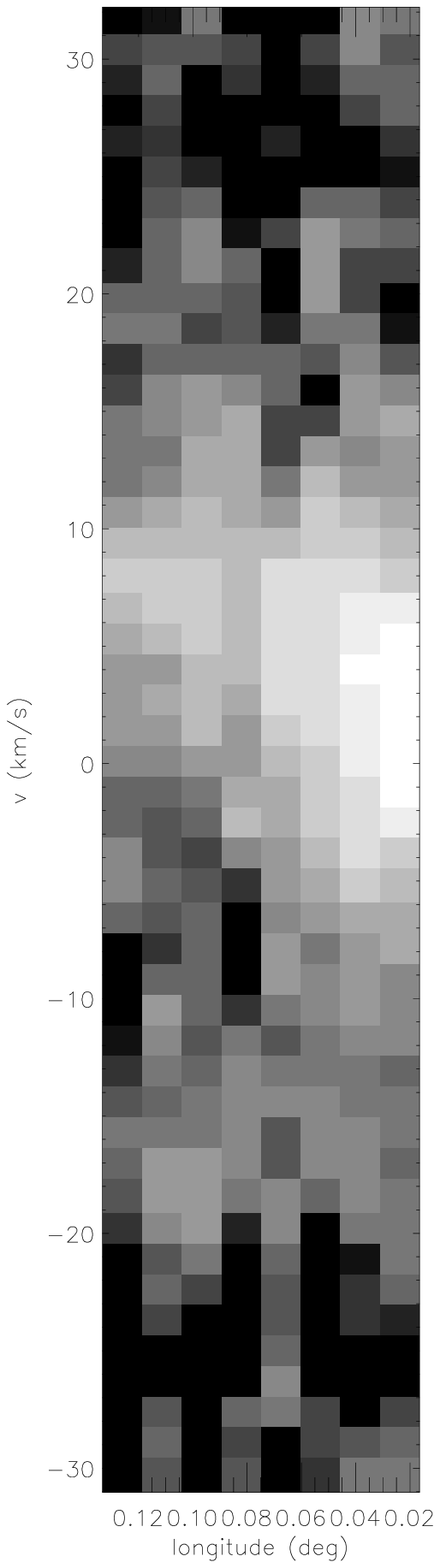}
\hfil
\includegraphics[scale=0.6]{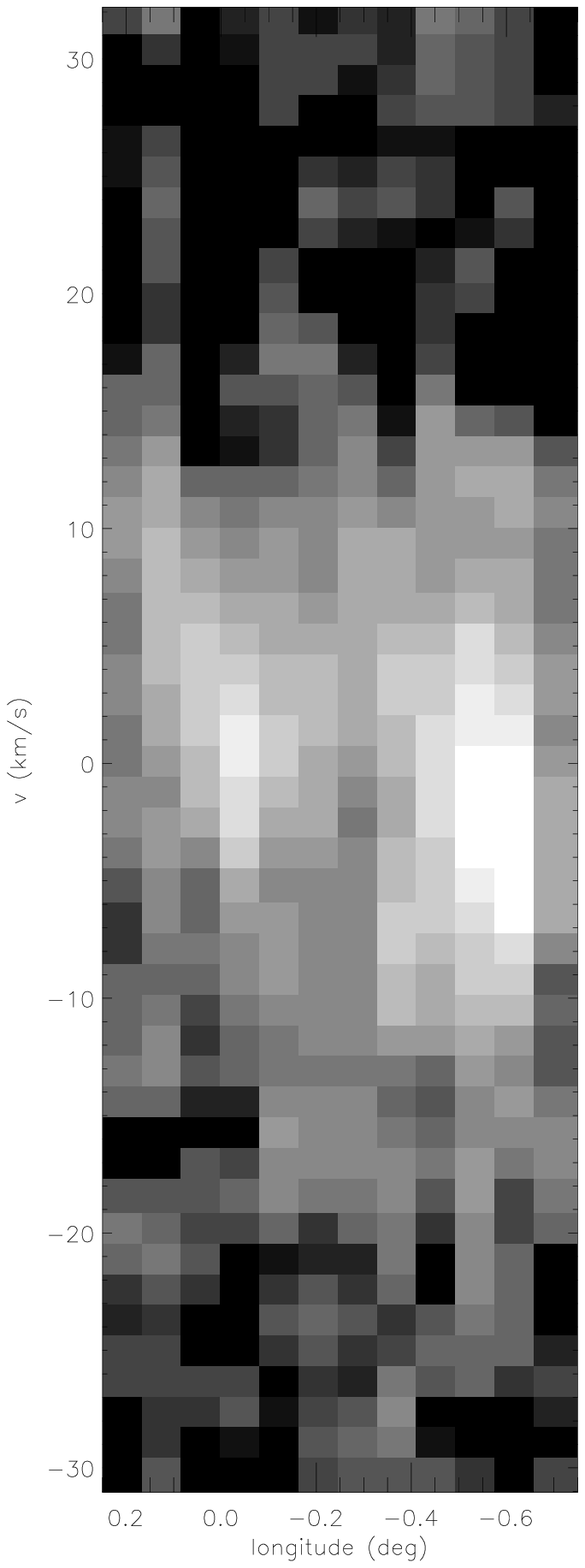}
\hfil
\includegraphics[scale=0.6]{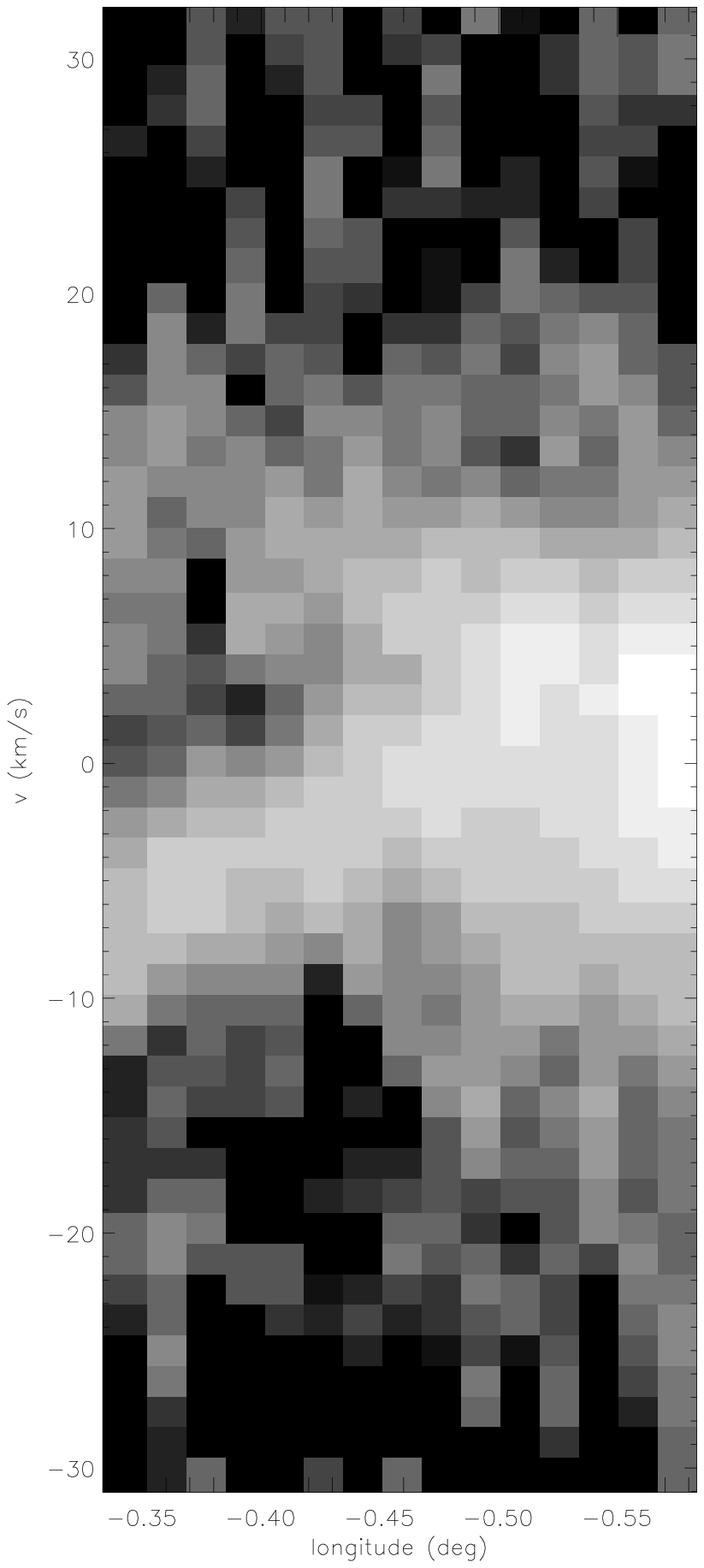}
\caption{$l$--$v$ diagrams of the average H$\alpha$ radio line brightness for the GCL3 strip (\emph{left}), $b=0\ddeg45$ strip (\emph{center}), and GCL4 strip (\emph{right}).  The GCL3 strip starts near the eastern peak of the recombination line emission of the GCL and goes toward the northeast.  The $b=0\ddeg45$ strip spans the entire east-west extent of the GCL.  The GCL4 strip starts near the western peak of the line emission in the GCL and goes toward the northeast.  Detailed descriptions of these strips of observations are given in Table \ref{gbtpointings}.  The right $l$--$v$ diagram does not account for the missing scan at $l=-0\ddeg48$, which gives errors in $l$ of order $1\damin4$ near that longitude. \label{lvimg}}
\end{figure}

Figure \ref{lvimg} shows an alternative way of visualizing the recombination line velocity structure using $l$--$v$ diagrams.  Three strips of observations show the change in the line velocity as a function of Galactic longitude.  These plots show clearly how the brightest emission is always within 20 \kms\ the rest velocity, and that the lines are typically about 10 \kms\ wide.  The plots also help show how the line profile has multiple components inside the GCL, for $l=-0\ddeg2$ to $-0\ddeg45$.  In particular, the GCL4 $l$--$v$ diagram shows two diagonal structures near $l=0\ddeg4$, which indicate that there are two narrow components to the line profile with velocities approaching $\pm10$ \kms\ toward the eastern end of the strip.

\begin{figure}[tbp]
\includegraphics[scale=0.15]{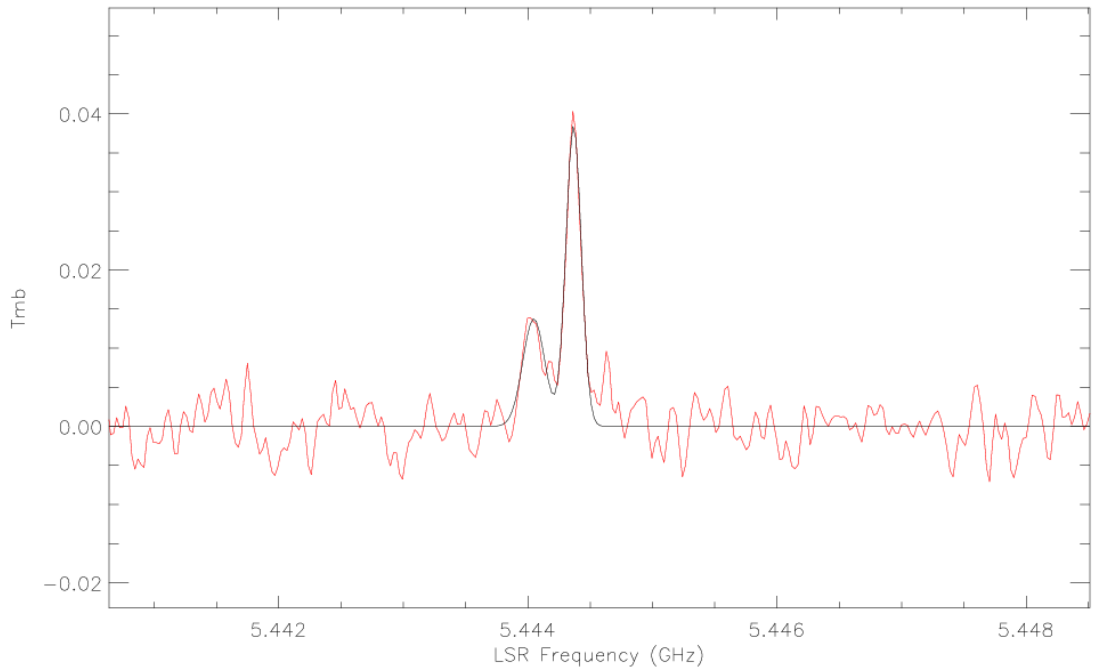}
\caption{Plot of the average H$\alpha$ spectrum for three scans inside the GCL, on the eastern edge of the GCL4 strip.  The profile requires two narrow Gaussians for an adequate fit, with velocities of roughly +14 and --6 \kms. The unusually narrow lines suggest that this gas is associated with the GCL \label{polcen2}}
\end{figure}

To compare the emission at the edges to the center, an average of the H$106-113\alpha$ spectra for scans at the edges of the GCL3 and GCL4 strips was made.  Figure \ref{polcen2} shows how the average of three scans from the eastern edge of GCL4 confirms the visual impression from the $l$--$v$ diagram:  the best-fit spectrum has two distinct components with $(T_l, v, \Delta v)=(13.8\pm1.4\ \rm{mK}, 13.0\pm1.6$ \kms$, 12.5\pm1.5$ \kms$)$ and $(38.5\pm1.6\ \rm{mK}, -6.4\pm0.2$ \kms$, 8.8\pm0.4$ \kms$)$.  A similar average over the four easternmost scans of the GCL3 strip (with $l=0\ddeg083-0\ddeg133$) shows two Gaussian components to the average profile.  The two components have $(T_l, v, \Delta v)=(26.9\pm1.0\ \rm{mK}, 7.6\pm0.3$ \kms$, 17.2\pm0.8$ \kms$)$ and $(11.1\pm 1.4\ \rm{mK}, -18.3\pm0.5$ \kms$, 8.6\pm1.3$ \kms$)$.  Both components are narrow and brighter than the background, so the emission is likely to be associated with the GCL.

Similar to the HCRO observations, the GBT observations find no emission with velocities $|v|>20$ \kms.  A deep spectrum was made by averaging observations at the peak GCL line emission, called GCL3 and GCL4 in Table \ref{gbtpointings};  the total on-source integration time for this deep spectrum was 27 minutes.  The rms noise in the baseline level of the deep GBT spectrum gives an upper limit to the peak line brightness of about 1 mK for $v=\pm500$ \kms\ and 2 mK for $v=\pm1500$ \kms  The peak line brightness limits are 1\% and 2\% of the peak at $v\sim0$ \kms, respectively.

\subsubsection{Line Widths and Line Ratios}
\label{gbt:lines}
To better constrain the gas conditions, we studied several recombination lines in the deep integration toward the peak GCL line emission.  The relatively small velocities observed in these integrations makes it possible to average them together.  The two 3$\times$3, Nyquist-sampled patterns near the brightest regions of the GCL were found to have not only similar line velocities and line widths, but also similar line ratios, suggesting that these regions have similar physical conditions.

\begin{figure}[tbp]
\begin{center}
\includegraphics[scale=0.15]{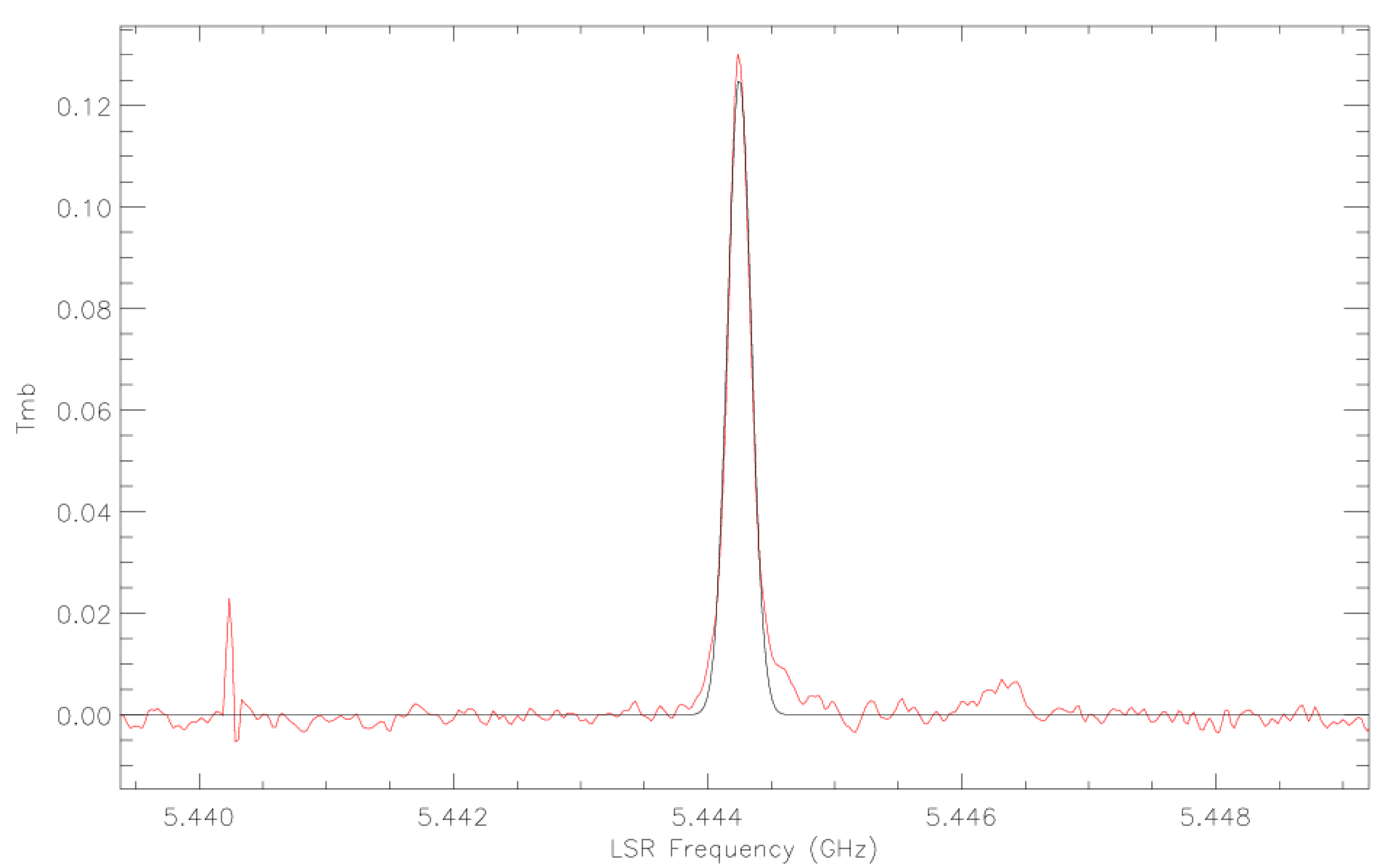}
\hfil
\includegraphics[scale=0.15]{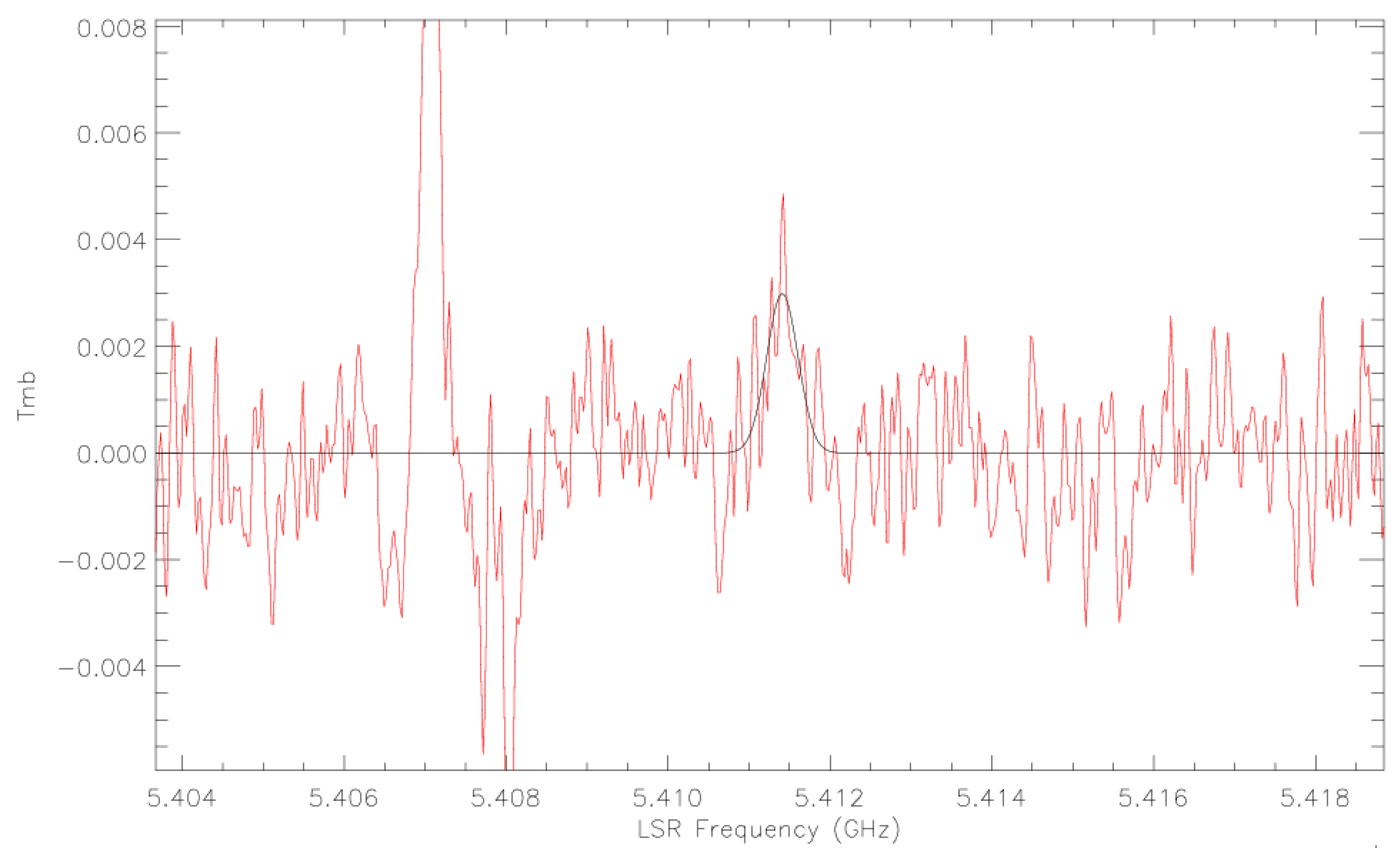}
\end{center}
\caption{\emph{Top}: Average H$\alpha$ spectrum for eight transitions with $n=$106 to 113 in the deep average spectrum of the GCL.  The best-fit Gaussian line described in Table \ref{lines} is overlaid.  The lump near $\nu=5.446$ GHz is the smeared average of the nearest He$\alpha$ transitions.   \emph{Bottom}:  Same as above, but for eight H$\epsilon$ transitions with $n=$180 to 191.  The brightest line, at $\nu=5.407$ GHz, is H108$\alpha$, which is about 4 MHz from H183$\epsilon$.  Note that the frequencies of the spectrum shown here correspond to that of the highest-frequency transition observed, but that all velocities and line properties are calculated by assuming an idealized ``mean'' of all transitions with a given $\Delta n$. \label{lineplots}}
\end{figure}

Table \ref{lines} lists the line measurements and Figure \ref{lineplots} shows the highest and lowest significance fits in the deep spectrum toward the GCL.  A Gaussian line is fit to the average line profile for all transitions of a given $\Delta n$ (e.g., $\Delta n=1$ for ``$\alpha$''), which includes 7 or 8 transitions in the range given in Table \ref{lines}.  Generally, these averages over several $n$ values are treated like a single transition with $n$ and $\nu\approx4.95$ GHz, or the mean of the transitions averaged.

One of the most surprising characteristics of the lines shown in Table \ref{lines} is their unusually small widths.  The average H$106-113\alpha$, H$134-142\beta$, H$152-162\gamma$, and H$167-178\delta$ transitions have widths that are similar to each other and have an error-weighted mean of $\Delta v=13.5\pm0.2$ \kms.  The average H$180-191\epsilon$ line is significantly wider and is discussed in more detail below.  Typical \hii\ regions have much larger line widths, with $\Delta v\approx20$ \kms\ \citep{s83,a96}.  A hydrogen recombination line width of $13.5\pm0.2$ \kms\ is equivalent to a Doppler temperature of $3960\pm120$ K \citep{r92}, which is also significantly less than values seen in typical \hii\ regions ($T_D\approx5000-10000$K).  The Doppler temperature represents the combined effects of gas motion and thermal broadening, so it is an upper limit on the electron temperature.  The line width is roughly constant for transitions from H$106-113\alpha$ to H$167-178\delta$, which represent transitions with mean $n$ ranging from 109.5 to 172.5.

The narrow line widths with central velocities near 0 \kms\ are highly suggestive of stimulated emission by ionized gas along the line of sight to the GC region \citep{c90,a97}.  The present observations can test for the possibility of stimulated line emission by comparing the line ratios of transitions with different $n$.  All transitions are observed with a similar beam size, which means that they are probing the same volume of space and can be meaningfully compared \citep{d70}.  In this case, the integrated line ratio between transitions to states $n$ and $m$ takes on the form
\begin{equation}
R \approx R_{\rm{LTE}} \frac{b_n}{b_m} \frac{T_e - T_c \beta_{n}}{T_e - T_c \beta_{m}},
\end{equation}
\noindent where $R_{\rm{LTE}}= n^2 f_{n,n+\Delta n}/m^2 f_{m,m+\Delta m}$, $b$ and $\beta$ are the departure coefficients for non-LTE effects, and $T_e$ and $T_c$ are the electron and background continuum brightness temperatures \citep{d70,b72,r92}.  The departure coefficients vary with $n_e$, $T_e$, and $n$, and have been calculated for many transitions under a wide range of gas conditions \citep{d72,s79}.  To estimate non-LTE effects, we calculated departure coefficients for $T_e=20-10000$ K and $n_e=1-1000$ cm$^{-3}$.  The lowest temperature and density used in the calculations correspond to the conditions that may cause stimulated emission observed toward the GC region at 1.4 GHz \citep{l73}.  The LTE line ratios depend on the oscillator strengths for the transitions, which have relative values known to high precision \citep{d69}.

Table \ref{lineratios} shows the integrated line ratios and the predictions for LTE and non-LTE conditions.  The ratios show excellent agreement with LTE for all transitions observed.  The predicted line ratios for other models do not agree with the observed line ratios.

One model, with $T_e=1000$ K and $n_e=1000$ cm$^{-3}$, predicts line ratios similar to the observed and LTE values.  The amount of stimulated emission can be calculated as
\begin{equation}
T_l/T_l^{\rm{LTE}} \approx b_n [1 - \beta_n (T_c/T_e)],
\end{equation}
\noindent where $T_l$ is the line brightness, and the background continuum brightness is $T_c\approx1-2$ K \citep{gcsurvey_gbt}.  For $T_e=1000$ K and $n_e=1000$ cm$^{-3}$, the amplification of the line is only a few percent.  A much stronger background continuum is required to cause significant stimulated emission, with 10\% amplification for $T_c\sim25$ K;  this is more than 10 times brighter than the observed continuum in the region.  Thus, the line ratios severely constrain the contribution of stimulated emission to the recombination line emission for the brightest emission observed toward the GCL.

There are two other points that suggest that the recombination line emission is not stimulated.  First, if background radio continuum was stimulating the recombination line emission, the morphology of the line emission would closely follow the background continuum emission \citep{l73,c90}.  The HCRO and GBT observations show that the peak line emission, particularly in the eastern half of the GCL, is significantly offset from the peak continuum emission.  The similar line width, line velocity, and line ratios toward the eastern and western recombination line peaks suggest that the entire structure emits by a similar mechanism.  Second, the width of the average H$180-191\epsilon$ line is significantly larger than that of the other lines, which is inconsistent with the idea that it is stimulated.  However, the H$189-191\epsilon$ line represents the highest $n$ states, where stimulated emission is expected to be strongest.  These points strengthen the case for an LTE origin for the recombination line emission, which implies that the narrow line widths are related to intrinsic properties of the ionized gas.

Using the LTE assumption, the line properties give useful constraints on the intrinsic gas properties.  As mentioned before, the width of the average H$180-191\epsilon$ line is significantly larger than the other transitions.  Widths of lines that are not stimulated are generally dominated by thermal and Doppler broadening, although these effects are not expected to change between the different transitions observed here.  The most common $n$-dependent broadening effect is collisional broadening (a.k.a. ``impact'' broadening), in which inelastic electron collisions preferentially broaden lines with large $n$ \citep{l78,r92}.  Expressing the total line width as the quadrature sum of all broadening effects gives:
\small
\begin{equation}
\frac{\Delta v_{\rm{tot}}}{\rm{km\ s}^{-1}} = \sqrt{\left[\frac{4.31}{\Delta n} \left(\frac{n}{100}\right)^{7.4} \frac{n_e}{10^4\ \rm{cm}^{-3}} \left(\frac{T_e}{10^4\ \rm{K}}\right)^{-0.1}\right]^2 + \Delta v_d^2},
\label{widtheqn}
\end{equation}
\normalsize
\noindent where $\Delta v_d$ is the Doppler line width. \footnote{Technically, collisional broadening takes the shape of a Lorentz distribution, which, when convolved with a Gaussian, produces a Voigt profile \citep{r92}.  However, at the low signal to noise ratio observed for the H$180-191\epsilon$ transition, no line wings are apparent and a Gaussian is a close approximation to the expected line shape.}

\begin{figure}[tbp]
\begin{center}
\includegraphics[scale=0.7]{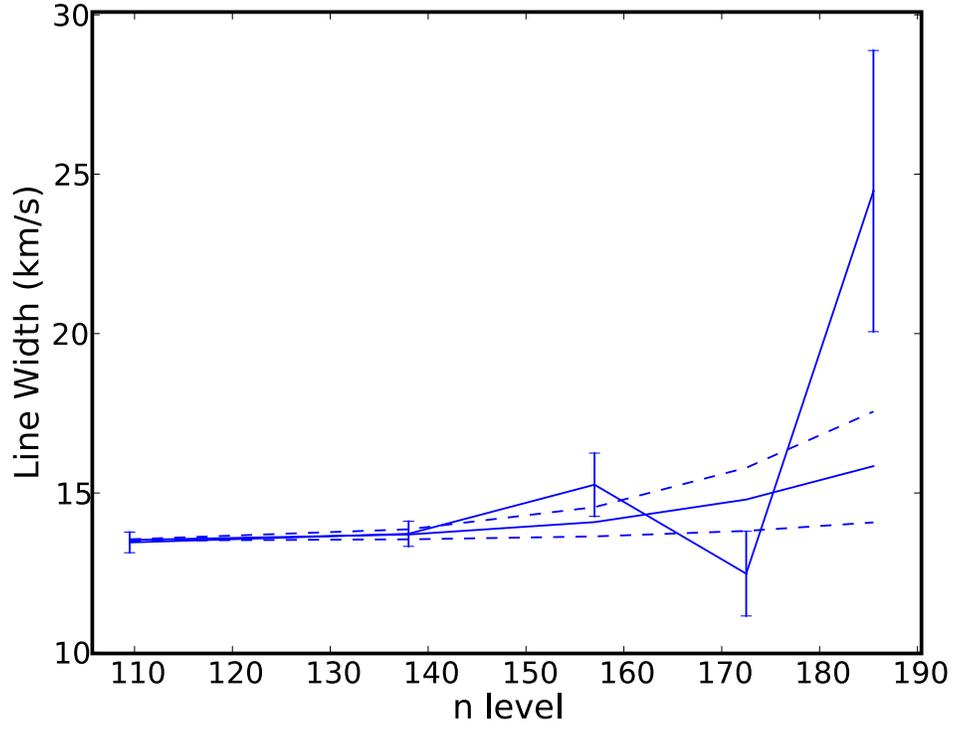}
\end{center}
\caption{Plot of the distribution of hydrogen line widths observed by the GBT as a function of electronic state $n$.  The solid line shows the best-fit model to the widths, considering the effects of collisional broadening; the dotted lines show the $\pm1 \sigma$ fits.  The normalization of the curves corresponds to an electron density $n_e = 910^{+310}_{-450}$ cm$^{-3}$, assuming $T_e=3960$ K. \label{widths}}
\end{figure}

The strong dependence of collisional broadening on $n$ can be used to constrain $n_e$, since $\Delta v_d$ is known and the dependence on $T_e$ is weak.  Equation \ref{widtheqn} was fit to the distribution of Hydrogen line widths assuming $\Delta v_d=13.5$ \kms, which is the error-weighted mean of the H$106-113\alpha$ to H$167-178\delta$ line widths \footnote{Allowing $\Delta v_d$ to vary did not significantly change the fit values or its quality.} and $T_e = T_d = 3960$ K.  Figure \ref{widths} shows the best-fit line width model with $n_e=950^{+240}_{-310}$ cm$^{-3}$ (1$\sigma$).  The fit quality is relatively poor, with $\chi^2/\nu = 14.4/4$, which suggests that the errors may be underestimated somewhat.  For a more realistic fit, we model additional uncertainty in the line widths as equal to the statistical uncertainty for the H$180-191\epsilon$ line and scaled by the line peak signal to noise ratio;  this uncertainty is consistent with variation in the fit line width with changes in the baseline fit.  Adding this line width uncertainty in quadrature to the statistical error, the best-fit electron density is $n_e=910^{+310}_{-450}$ cm$^{-3}$ with a more realistic $\chi_\nu^2=8.4/4$.

The He$106-113\alpha$ line was detected in the deep spectrum with 18$\sigma$ significance.  The ionized helium abundance is measured by the integrated line ratios $Y^+=I_{\rm{He}\alpha}/I_{\rm{H}\alpha}$.  In the GCL, this ratio is $0.075\pm0.007$, similar to the solar value \citep{l74,r92}.

Typically, the line-to-continuum ratio for a radio recombination line is a useful way to measure the electron temperature of the emitting region.  The LTE electron temperature is:
\begin{equation}
T_e^{\rm{LTE}}=\left[6943 \frac{\nu}{\rm{GHz}}^{1.1} \frac{T_c}{T_l \Delta v} \frac{1}{1+Y^+}\right]^{0.87} \rm{K},
\end{equation}
\noindent which is related to the actual electron temperature by:
\begin{equation}
T_e^{\rm{LTE}} = T_e \left[b_n \left(1 - \frac{\beta_n \tau_c}{2}\right)\right]^{-0.87},
\end{equation}
\noindent where $\tau_c$ is the optical depth of the background continuum emission.  However, much of the continuum emission in the GCL is nonthermal \citep{gcsurvey_gbt}, so the line-to-continuum ratio is an upper limit on the electron temperature.  For the deep spectrum, the continuum flux was estimated from the 5 GHz map described in \citet{gcsurvey_gbt} to be $T_c\approx0.8$ K, which gives $T_l \Delta v/T_c=2.0$ \kms, for the averaged H$106-113\alpha$ line.
This line-to-continuum ratio gives to an upper limit on the LTE electron temperature of $T_e^{\rm{LTE}}\lesssim5220$ K, which is consistent with the upper limit on $T_e$ from the narrow line widths.  For the largest non-LTE effects that are consistent with the observed line ratios ($n_e=100$\ cm$^{-3}$ and $T_e=1000$\ K), $T_e = (0.855)^{0.87} T_e^{\rm{LTE}}\lesssim4550$ K, but for higher densities and temperatures, $T_e$ is within a few percent of $T_e^{\rm{LTE}}$.  Alternatively, we can use the observed line width to constrain the electron temperature.  The upper limit to the electron temperature gives a lower limit on the line-to-continuum ratio of $T_l \Delta v/T_c>2.75$ \kms\ and an upper limit on the thermal continuum of 0.6 K.  Thus, at most about 70\% of the radio continuum toward the recombination line peaks of the GCL is thermal.

The upper limit to the electron temperature constrains the emission measure.  Assuming $\tau_L\ll1$ and LTE for the average H$106-113\alpha$ transition, the continuum emission measure is:
\begin{equation}
\rm{EM}_c = 8.5\times10^{-3} (1+Y^+) \Delta v T_l T_e^{3/2} e^{-X_n} \rm{pc}\ \rm{cm}^{-6},
\end{equation}
\noindent where $\Delta v$ is in units of \kms and $e^{-X_n}\approx1$, for $T_e\gtrsim1000$ K \citep{d70}.  Using the average H$106-113\alpha$ line parameters for the deep spectrum gives EM$_c\approx3850 (T_e/3960\ \rm{K})^{3/2}$ pc cm$^{-6}$.  The eastern and western peak line brightnesses are somewhat different, giving EM$_c\approx3080 (T_e/3960\ \rm{K})^{3/2}$ and $4570 (T_e/3960\ \rm{K})^{3/2}$ pc cm$^{-6}$ for the east and west, respectively.

\section{Discussion}
\label{recomb_dis}

\subsection{The Structure of the Ionized Gas}
\label{disc:morph}
The GBT and HCRO observations have found radio recombination line emission throughout the radio continuum emission of the GCL.  The strongest emission is found along an eastern and western ridge, but significant emission is found between these ridges.  All of this emission has an unusually narrow line width and LTE line ratios, suggesting that it is all a part of the same structure and distinct from background Galactic emission.

The morphology of the ionized gas suggests that it is associated with the radio continuum emission of the GCL.  The two ridges and cap of recombination line emission are parallel and adjacent to similar structures in radio continuum maps \citep{gcsurvey_gbt}.  If the radio recombination line and continuum emission are colocated, the line emission may be in the GC region, as has been argued for radio continuum GCL \citep{l89,b03}.  A more detailed argument is left for a future paper;  for the results shown below, we derive physical parameters for the ionized gas assuming a distance of 8 kpc.

The morphology of the line emission is consistent with a three-dimensional shell model.  In this model, the two ridges of emission are the shell edges, where the column density is largest.  \citet{b03} model the \emph{MSX} mid-IR emission from this region as a ``telescope dome'' shell specified by a height, radius, and shell thickness.  Assuming this geometry, the radius of the dome is roughly $0\ddeg3$ and the height is roughly 1\sdeg.  The GBT observations measured the ridge half-peak thickness to be about $0\ddeg2$;  simple simulations suggest that for a shell geometry, the true shell width is about half the apparent thickness, or $0\ddeg1$.  Assuming this shell is in the GC region, the dome radius is 40 pc, the total height is 140 pc, and the width is 15 pc.  The path length through the shell edge would be about 50 pc.  Combining this with the $EM$ measurement, gives an rms electron density, $\sqrt{\langle n_e^2\rangle}\approx8.8 (T_e/3960\ \rm{K})^{3/4}$ cm$^{-3}$.

The morphology of the GCL is reminiscent of ``Galactic worms'' \citep{k92}.  Worms are ionized cavities formed when supernovae and stellar winds from \hii\ regions blow out of the Galactic disk \citep{h96}.  A difference between the GCL and a worm is that the GCL has a cap at its top.  The GCL may be thought of as a frustrated worm, an ionized cavity that has not yet blown out of the Galactic disk.  A degree-scale structure like this would not have been detected by previous surveys for worms \citep{k92}.

\subsection{Properties of the Ionized Gas}
\label{disc:line}
It is critical to show that the recombination line emission is in LTE if we wish to derive intrinsic gas conditions.  Previous observations in the GC region found narrow recombination lines, particularly at lower frequencies, and concluded that the emission was most likely stimulated \citep{l73,p75,a97}.  However, there are several characteristics of the line emission observed here that show that the gas is in LTE, including well-constrained line ratios consistent with LTE, the lack of coincidence between background continuum brightness and line brightness, and evidence for collisional broadening in the line widths.

Assuming the emission is not stimulated, the Doppler widths of these lines provide a strict upper limit to the electron temperature of $3960\pm120$ K.  This limit to the electron temperature is among the lowest values observed in the Galaxy, although radio recombination line observations of other \hii\ regions have found electron and Doppler temperatures as low as 4000 K \citep[][M. Goss 2006, private communication]{s83,a96}.  Some of the best-fit line widths discussed in this work are as low as 9 \kms, which corresponds to a Doppler temperature of $\sim$1800 K, much smaller than observed elsewhere in the Galaxy.

The dominant effect in determining the electron temperature of ionized gas is the cooling efficiency;  high metal abundance allows ionized gas to cool quickly \citep{m85}.  Studies of recombination line emission from \hii\ regions have used this effect to establish that there is a metallicity gradient with distance from the GC \citep{s83,a96}.  The gradient is seen as a trend in $T_e$, with coolest and most metal-rich gas in the GC region.  The relation observed by \citet{a96} predicts $T_e\approx5500$ K in the GC region, while \citet{s83} predicts $T_e\approx3100$ K.  Regardless of the exact value, these relations suggest that the ionized gas in the GCL is located near the GC, supporting the morphological connection.

The broadening of the line width with increasing electronic state, $n$, is best explained by collisional broadening.  The best-fit model to the line widths gives $n_e=910^{+310}_{-450}$ cm$^{-3}$.  The best-fit density is consistent with the lack of stimulated emission, which would change the line ratios for $n_e\lesssim100$ cm$^{-3}$.  Using the rms and true electron densities, we calculate the volume filling factor for the deep spectrum as $f=\langle n_e^2\rangle/n_e^2=(9^{+27}_{-4})\times10^{-5} (T_e/3960\ \rm{K})^{1.3} (50\ {\rm{pc}}/L)$ cm$^{-3}$, where $L$ is the path length through the shell, assuming a distance of 8 kpc.  Since the emission measure scales as $\langle n_e^2\rangle$, the derived filling factor applies to the densest gas along the line of sight.  This filling factor is much smaller than that seen in the Galactic disk \citep[$f\sim10^{-2}$;][]{h96}, but comparable to that seen in Galactic outflows \citep[$f\sim10^{-3}-10^{-4}$;][]{h90}.

The thermal pressure implied by the constraint on $T_e$ and measurement of $n_e$ is $P/k=7.2\times10^6 (T_e/3960\ \rm{K})$ K cm$^{-3}$.  This pressure is about 100 times larger than the total gas pressure near the Sun \citep{bl87}, but it is not unusual for the GC region \citep{s92,ma04,k96,m04}.  This suggests that the ionized gas is in the GC region and in equilibrium with its molecular and hot, x-ray-emitting gas.

\subsection{Mass and Ionization of the Gas}
\label{derived}
Assuming a shell geometry and GC distance to the recombination line emission allows us to estimate other physical properties of the gas.  For ionized gas in LTE, the mass is parameterized as:
\small
\begin{equation}
M = 0.419 (T_e/10^4\ {\rm{K}})^{0.175} (S_{5 {\rm{GHz}}}/{\rm{Jy}})^{0.5} (D/{\rm{kpc}})^{2.5} (\theta_G/{\rm{arcmin}})^{1.5},
\end{equation}
\normalsize
\noindent where $D$ is the distance to the gas, $\theta_{\rm{G}}$ is the angular size of the emission, and the geometry is assumed to have a cylidrical shape \citep{p75,p78}.  Since the gas is in LTE, the line-to-continuum ratio is $T_l \Delta v/T_c=2.75 (T_e/3960\ \rm{K})^{-0.87}$ \kms, as shown in \S\ \ref{gbt:lines}.  The integrated line intensity over the ionized shell of gas in the HCRO observations is about $1.2\ \rm{K} * 13.5 $ \kms$ = 16 $ K \kms, for $b\ge0\ddeg2$ and subtracting 10\% for the background.  This gives a 5 GHz continuum flux of 50 Jy and a ionized mass $M=2\times10^5 (T_e/3960\ \rm{K})^{0.61}$ \msol\ in the shell.  Uncertainties in the mass estimate are probably dominated by incomplete sampling of the HCRO survey and are about 20\%.

The mean electron density constrains the number of Lyman continuum photons required to ionize the gas.  The Lyman continuum photon flux is $N_{\rm{Ly}} = V n_e n_{\rm{H}} \alpha^{(2)}$, where $V$ is the volume of the emitting region and $\alpha^{(2)}\approx6.6\times10^{-13}$ cm$^3$ s$^{-1}$, for $T_e=3\times10^3$\ K \citep{s88}.  Using the integrated thermal continuum flux density, we find that the gas requires an ionizing flux $N_{Ly} = 7\times10^{49} (T_e/3960\ \rm{K})^{1.22})$ s$^{-1}$.  

The estimated ionizing flux is about seven times the Lyman continuum flux of an O7 V star of about $10^{49}$ s$^{-1}$ \citep{sm02}.  Assuming it is a shell in the GC region, the source of ionizing photons would most likely be stars in the plane.  Based on geometry, at most half of the stellar flux would reach the shell, so the required ionizing flux is equivalent to more than 14 07V-type stars, or $1.4\times10^{50}$ s$^{-1}$.  The Arches, Quintuplet, and Central star clusters, all located inside the GCL in projection, have measured Lyman continuum fluxes of $10^{51.0}$, $10^{50.9}$, and $10^{50.5}$ s$^{-1}$ \citep{f99}.  Even if extinction is large, these clusters could easily ionize the gas associated with the GCL.

\subsection{Kinematics}
\label{kinematics}
The observations confirm previous observations that found line velocities near zero toward the GCL \citep{l73}.  However, the GBT $l$--$v$ diagrams show that the line velocities tend to be much more complex, particularly between the bright ridges of emission, from $l=0\ddeg0$ to $-0\ddeg5$.  Unusually narrow lines are brightest near the GCL and have physical conditions most consistent with the GC region.  This narrow line observed across the GCL at $b=0\ddeg45$ shows a velocity gradient similar to Galactic rotation, indicating that the gas may be connected to the disk.


The low velocity of the ionized gas has been used to argue that it is in the foreground to the GC region \citep{p75}.  However, recent work has shown that a significant fraction of gas in the central few hundred parsecs has velocities near zero.  \citet{o05} studied $H_3^+$ absorption lines toward the GC region;  these lines are excited at temperatures and densities found in the central few hundred parsecs and have only been observed toward the GC region.  They found that roughly 1/3 of $H_3^+$ column density has velocities near zero.  Thus, low velocity gas is quite common in the GC region.

\citet{s96} argued that the GCL is associated with molecular gas rotating about the GC at $\pm\sim100$ \kms.  While that association relies on a positional coincidence in a complex region, it is worth considering how it would relate to the ionized gas.  In particular, how can the large difference in velocity of these two components be explained?  The motion of this molecular gas is best explained by the gas dynamics in a barred gravitational potential \citep{b91}.  Under this model, the gas possibly associated with the GCL is either in or transitioning to the inner ``x2'' orbits \citep{c77}.  In light of this, the low velocity of the ionized gas in the GCL may be tied to the shocked molecular gas on these transitioning orbits.

Alternatively, the low velocity of the ionized gas may indicate that it as been decelerated by an ambient magnetic field.  Equating gas ram pressure to a magnetic pressure allows us to estimate the magnetic field required to stop the gas on a dynamic time scale.  For the molecular gas associated with the GCL by \citet{s96}, these pressures are equal for $B=1$ mG.  The possibility of large-scale, mG-strength fields in the GC region is being actively debated \citep{y87,l05,b06};  a weaker field would decelerate more slowly, but could ultimately have the same effect.  

\section{Conclusions}
\label{conc}
We have presented analysis of new and archival radio recombination line observations toward the GCL, a radio continuum shell believed to be evidence of a mass outflow from our GC region.  The observations have found that the radio line emission has a morphology strikingly similar to the radio continuum, strongly arguing that they are associated.  The line and continuum emission appears as a limb-brightened shell.  A simple shell model for the structure is best fit with a radius of $0\ddeg3$, a height 1\sdeg, and a shell width $0\ddeg1$.

Recombination lines are detected in averaged profiles covering transitions from H106$\alpha$ up to H191$\epsilon$.  Diagnostics derived from the detected lines show that the emission is not stimulated, allowing us to constrain the electron temperature and gas density from the line widths.  The electron temperature is unusually low, with an upper limit of roughly 4000 K.  This temperature limit, and the gas pressure, are consistent with conditions in the GC region.

Assuming that the gas is organized as a shell and is in the GC region, we derive a mass of $M=2\times10^5 (T_e/3960\ \rm{K})^{0.61}$ \msol.  This mass is similar to that observed in dwarf starburst outflows \citep{v05}.  The filling factor of the ionized gas is also similar to that seen in starburst outflows.  In the canonical model for such outflows, the emission comes from the ionized surface of clumps of molecular gas entrained in the outflow \citep{h90}.  

Modest observations with radio interferometers currently in development could greatly expand on the results presented here.  A simple estimate using the filling factor and the size of our deep integration shows that these clumps could be resolved at size scales of $\sim$9\arcsec.  Deeper observations of more transitions would strengthen the significance of collisional broadening and detect spatial structure in the electron density and gas pressure.  Finally, high-spectral resolution observations would map the line velocities and help us understand its dynamics.

\acknowledgements{We thank Miller Goss for an enlightening discussion and the anonymous referee for useful comments.}

{\it Facilities:} \facility{GBT (), HCRO ()}

\clearpage

\begin{deluxetable}{lcclcc}
\tablecaption{GBT Radio Recombination Line Pointings \label{gbtpointings}}
\tablewidth{0pt}
\tablehead{
\colhead{Name} & \colhead{$l$} & \colhead{$b$} & \colhead{Map type} & \colhead{Exposure} & \colhead{Spacing}\\
\colhead{} & \colhead{(deg)} & \colhead{(deg)} & \colhead{} & \colhead{(s)} & \colhead{(arcmin)} \\
}
\startdata
GCL3 & 0.0 & 0.5 & 3x3 grid & 90 & 1 \\
GCL4 & 359.4 & 0.5 & 3x3 grid & 90 & 1 \\
GCL3 strip & +0.016--0.133 & 0.516--0.633 & diagonal strip & 60 & 1.4 \\
GCL4 strip & 359.416--359.666\tablenotemark{a} & 0.516--0.766 & diagonal strip & 60 & 1.4 \\
$b=0.45$ strip & 0.25--359.25 & 0.45 & horizontal strip & 60 & 5.4 \\
GCL1 & 0.3 & 0.5 & single point & 60 & -- \\
GCL7 & 359.1 & 0.8 & single point & 60 & -- \\
\enddata
\tablenotetext{a}{One pointing, near $l=-0\ddeg48$, was accidentally skipped.}
\end{deluxetable}

\begin{deluxetable}{lccccc}
\tablecaption{Detected Lines and their Properties in Deep Integration Toward GCL Recombination Line Peaks \label{lines}}
\tabletypesize{\scriptsize}
\tablewidth{0pt}
\tablehead{
\colhead{Transition} & \colhead{$T_l$} & \colhead{$v_{\rm{LSR}}$} & \colhead{$\Delta v$} & \colhead{$I^{\rm{obs}}$\tablenotemark{a}} & \colhead{$R^{\rm{obs}}_\alpha$\tablenotemark{b}} \\
\colhead{} & \colhead{(K)} & \colhead{(\kms)} & \colhead{(\kms)} & \colhead{(K \kms)} & \colhead{} \\
}
\startdata
H$106-113\alpha$ & $0.1257\pm0.0024$ & $0.71\pm0.13$ & $13.45\pm0.30$ & $1.802\pm0.052$ & 1 \\
He$106-113\alpha$ & $0.0110\pm0.0006$ & $0.97\pm0.32$ & $11.50\pm0.74$ & $0.135\pm0.012$ & $0.075\pm0.007$ \\
H$134-142\beta$ & $0.0339\pm0.0008$ & $0.75\pm0.15$ & $13.72\pm0.36$ & $0.495\pm0.017$ & $0.275\pm0.012$ \\
H$152-162\gamma$ & $0.0144\pm0.0007$ & $0.36\pm0.35$ & $15.26\pm0.83$ & $0.233\pm0.017$ & $0.130\pm0.010$ \\
H$167-178\delta$ & $0.0091\pm0.0006$ & $1.98\pm0.40$ & $12.47\pm1.05$ & $0.120\pm0.013$ & $0.067\pm0.007$ \\
H$180-191\epsilon$ & $0.0030\pm0.0003$ & $1.67\pm1.34$ & $24.47\pm3.15$ & $0.073\pm0.012$ & $0.041\pm0.007$ \\
\enddata
\tablenotetext{a}{$I^{\rm{obs}} = \int_{\rm{line}} T_{\rm{mb}} dv \approx 1.065 T_l \Delta v$}
\tablenotetext{b}{$R^{\rm{obs}}_\alpha = I^{\rm{obs}}_{x}/I^{\rm{obs}}_\alpha$}
\end{deluxetable}

\begin{deluxetable}{lc|c|ccccccccc}
\tablecaption{Comparing Line Ratios to LTE and Stimulated Emission Models \label{lineratios}}
\tabletypesize{\scriptsize}
\tablewidth{0pt}
\tablehead{
\colhead{Transition} & \colhead{$R^{\rm{obs}}_\alpha$} & \colhead{$R^{\rm{LTE}}_\alpha$} & \colhead{$R^{\rm{20,1}}_\alpha$} & \colhead{$R^{\rm{100,10}}_\alpha$} & \colhead{$R^{\rm{100,100}}_\alpha$} & \colhead{$R^{\rm{100,1000}}_\alpha$} & \colhead{$R^{\rm{1000,100}}_\alpha$} & \colhead{$R^{\rm{1000,1000}}_\alpha$} & \colhead{$R^{\rm{10000,10}}_\alpha$} & \colhead{$R^{\rm{10000,100}}_\alpha$} & \colhead{$R^{\rm{10000,1000}}_\alpha$} \\
}
\startdata
H$106-113\alpha$ & 1               & 1      & 1     & 1     & 1     & 1     & 1     & 1     &  1    & 1     & 1 \\
H$134-142\beta$ & 0.275$\pm$0.012   & 0.279  & 0.322 & 0.381 & 0.300 & 0.295 & 0.281 & 0.275 & 0.417 & 0.242 & 0.245 \\
H$152-162\gamma$ & 0.130$\pm$0.010  & 0.127  & 0.224 & 0.198 & 0.137 & 0.133 & 0.123 & 0.124 & 0.193 & 0.098 & 0.103 \\
H$167-178\delta$ & 0.067$\pm$0.007  & 0.073  & 0.146 & 0.118 & 0.083 & 0.077 & 0.071 & 0.072 & 0.107 & 0.053 & 0.058 \\
H$180-191\epsilon$ & 0.041$\pm$0.007 & 0.047 & 0.112 & 0.077 & 0.055 & 0.049 & 0.053 & 0.048 & 0.055 & 0.051 & 0.047 \\
\enddata
\tablecomments{The line ratios for LTE conditions correspond to the ``mean'' transition of the range of $n$ values shown.  Oscillator strengths were taken at the mean $n$ or the average of the nearest two $n$ values, for fractional $n$.  Stimulated emission models labeled with ($T_e, n_e$) in K and cm$^{-3}$.  \citet{d72} and \citet{s79} were used for $T\leq100$ K and $>100$ K, respectively.}
\end{deluxetable}

\end{document}